\documentclass[aps,prl,floats,twocolumn,showpacs,amssymb]{revtex4}
\usepackage[dvips]{graphicx,color}
\begin{document}
%%%%%%%%%%%%%
\title{Negative differential conductance induced by spin-charge
  separation.}
%%%%%%%%%%%%%
\author{F. Cavaliere$^{1}$, A. Braggio$^{1}$, J. T.  Stockburger$^2$, M.
  Sassetti$^{1}$, and B. Kramer$^{3}$ \vspace{1mm}}
%%%%%%%%%%%%%%%
\affiliation{$^{1}$ Dipartimento di Fisica, INFM-Lamia, Universit\`{a} di Genova,
  Via Dodecaneso 33, 16146 Genova, Italy \\
  $^2$ II. Institut f\"ur Theoretische Physik, Universit\"at Stuttgart,
  Pfaffenwaldring 57, 70550 Stuttgart, Germany\\
  $^{3}$ I. Institut f\"ur Theoretische Physik, Universit\"at Hamburg,
  Jungiusstra\ss{}e 9, 20355 Hamburg, Germany\vspace{3mm}}
%%%%%%%%%%%%%%%%%%%%
\date{May 04, 2004}%%%%%%%%%%%%%%%%% with final changes by BK and MS
%%%%%%%%%%%%%%%%%%%%%%%%%%%%%%%%%%%%%%%%%%%%%%%%%%%%%%%%%%%%%%%%%%%%
\begin{abstract} Spin-charge states of correlated electrons in a
  one-dimensional quantum dot attached to interacting leads are studied in the
  non-linear transport regime. With non-symmetric tunnel barriers, regions of
  negative differential conductance induced by spin-charge separation are
  found. They are due to a correlation-induced trapping of higher-spin states
  without magnetic field, and associated with a strong increase in the
  fluctuations of the electron spin.
\end{abstract}
\pacs{73.63.Kv, 71.10.Pm, 73.22.Lp}
\maketitle
%%%%%%%%%%%%%%%%%%%%%%%%%%%%%%%%%%%%%%%%%%%%%%%%%%%%%%%%%%%%%%%%%%%%%%
In single electron transport \cite{ketal97}, the electron charge causes
the Coulomb blockade effect that stabilizes the electron number in a device.
Experiments show also that the electron spin can lead to important mesoscopic
transport effects. For instance, quantum dots in carbon nanotubes
\cite{lbp02,betal02,cn02} have revealed several non-equilibrium and coherent
spin processes, and spin-parity effects.

For {\rm few}-electron quantum dots, the spin blockade effect has been
predicted \cite{w95}. In one dimension (1D) this is indicated by a negative
differential conductance that occurs {\em only} when a state with maximum spin
value is occupied. Combining spin blockade with spin-polarized detection, the
electron spin in a 2D-lateral quantum dot was probed \cite{cetal02}. Also, in
a few-electron vertical quantum dot higher-spin states, not directly
accessible from the ground state, were detected via probing with
non-equilibrium voltage pulses \cite{fetal02}. When spin relaxation was
absent, strong fluctuations of spin and charge were observed. Motivated by
experiments on a quasi-1D quantum dot formed by two impurities in a quantum
wire \cite{aetal00}, we have calculated microscopically single electron
transport through a 1D quantum dot formed by two equal tunnel barriers in a
Tomonaga-Luttinger liquid. Signatures of non-Fermi liquid behavior and, in the
presence of a magnetic field, correlation-enhanced spin polarization were
found \cite{bsk01}. 
%HERE ADDED
Spin-charge separation has been observed \cite{aetal02} and analyzed
theoretically \cite{tetal02} in tunneling between parallel quantum
wires. In the {\em coherent} tunneling regime, the effects of
non-Fermi liquid correlations in 1D quantum dots have been
investigated non-perturbatively \cite{ng03,pg03}.

In the present work, we consider a 1D quantum dot in the incoherent
sequential tunneling region.
%END OF CHANGE 
We predict that states with higher total electron spin in a
non-magnetic 1D quantum dot containing an {\em arbitrary} number of
electrons can be dynamically stabilized by tuning the asymmetry of
tunnel barriers. The occupation of these states can lead, in the
sequential tunneling regime, to a negative differential conductance.
This is caused by the peculiar nature of the {\em non-Fermi liquid
  correlations} that determine the tunneling rates.  The new
phenomenon is due to {\em spin-charge separation} and is different
from the previously discussed negative conductance associated with
spin selection rules via Clebsch-Gordan coefficients that were
included in the tunneling rates {\em ad hoc} \cite{w95}. We study the
combined effects of asymmetry, electron correlations and relaxation.
We predict that stabilization of the higher-spin states is associated
with a strong increase of the spin fluctuations. Relaxation of these
states strongly affects the negative differential conductance, in
contrast to ``normal'' states associated with positive conductances.
Our results open a novel possibility of experimentally addressing
non-Fermi liquid behavior without performing often cumbersome analyses
of temperature dependences.

We start from the microscopic Hamiltonian $H=H_{\rm d}+H_{\rm l}+H_{\rm t}$,
which contains the dot ($H_{\rm d}$), the leads ($H_{\rm l}$), and a tunneling
term ($H_{\rm t}$). The dot is modeled as an interacting 1D system confined to
$|x|<d/2$, with excitations associated with {\em independent}
energy scales~\cite{bsk01}
\begin{equation}
  \label{eq:1}
H_{\rm d}(n,s,\rho,\sigma)=\frac{E_{\rho}}{2}(n-n_{\rm g})^2
  +\frac{E_{\sigma}}{2}s^2
  +\rho\varepsilon_{\rho}+\sigma\varepsilon_{\sigma}\,.
\end{equation}
Here, $n$ and $s$ represent the number of electrons, measured relative to the
number of charges $n_{\rm g}$ corresponding to a gate voltage $V_{\rm g}$, and
the $z$-component of the total spin in units of $\hbar/2$, respectively.
Energy scales $E_{\rho}$ and $E_{\sigma}$ are the charge and spin addition
energies. The last two terms do not change the total charge and spin but
describe intra-dot charge and spin density waves, where $\rho$ and $\sigma$
are related to bosonic creation and annihilation operators via $\rho = \sum_j
jb_{\rho,j}^\dagger b_{\rho,j}$ and $\sigma = \sum_j jb_{\sigma,j}^\dagger
b_{\sigma,j}$.

The energy parameters in Eq.~(\ref{eq:1}) reflect the microscopic electron
interaction parameterized by~\cite{voit}
\begin{equation}
  \label{eq:2}
g_{\rho}^{2}=\frac{1+V_{\rm x}}{1-V_{\rm x}+4V_{0}},
\quad   
  g_{\sigma}^{2}=\frac{1+V_{\rm x}}{1-V_{\rm x}}\,,
\end{equation}
with the exchange and Coulomb matrix elements $V_{\rm
  x}=\hat{V}(2k_{\rm F})/2\pi \hbar v_{\rm F}$ and
$V_{0}=\hat{V}(q=0)/2\pi \hbar v_{\rm F}$, respectively. In
experiment, the charging energy is not solely influenced by the
microscopic interaction in the quantum dot. Therefore, we treat it as
an independent parameter, typically $E_\rho \gg E_\sigma$. The spin
addition energy, $E_{\sigma}=\pi \hbar v_{\sigma}/2dg_{\sigma}$, is
due to the Pauli principle. The excitation energies of the neutral
charge and spin modes are $\varepsilon_{\nu}=\pi \hbar v_{\nu}/d$ with
charge and spin mode velocities $v_{\nu}=v_{\rm F}(1+V_{\rm
  x})/g_{\nu}$ ($\nu=\rho,\sigma$).  Without interaction,
$g_{\rho}=g_{\sigma}=1$, we have $E_{\sigma}\equiv E_{0}=\pi \hbar
v_{\rm F}/2d$ and $\varepsilon_{\rho}=\varepsilon_{\sigma}=2E_{0}$. We
assume the leads to be Luttinger liquids with Coulomb repulsion
$g_0\leq 1$, low-energy charge and spin mode dispersions
$\omega_{\rho}(q)=v_{\rm F}|q|/g_{0}$ and $\omega_{\sigma}(q)=v_{\rm
  F}|q|$. We assume $g_0\neq g_\rho$, since the
interaction can differ between dot and leads. Tunneling between the
leads and the dot at $x=\pm d/2$ is described by $H_{\rm t}$ with
amplitudes $\Delta_{\rm l}$ and $\Delta_{\rm r}$. We consider weak
tunneling \cite{bsk01}.

For not too low temperatures, $k_{\rm B}T\gg \delta E$, transport is
dominated by incoherent sequential tunneling and one can safely
neglect higher order coherent processes \cite{furu}.  Here, $\delta E$
represents the level broadening due to higher order contribution from
tunneling via virtual states, and is proportional to the rate $\gamma$
in Eq.~(\ref{eq:7}). The kinetic variables are then $n$ and $s$ {\em
  only} since $\rho$ and $\sigma$ relax towards thermal equilibrium
very efficiently due to mechanisms such as coupling to phonons and
spin-orbit interaction. The opposite limit, but for the spinless case,
has been considered in \cite{kinaret}. The relaxation of the spin $s$, which
is effected by the same processes as well as co-tunneling, is usually
much slower \cite{kn01}. Therefore, both tunneling and spin relaxation
have to be included in the Master equation. In order to be able to
resolve the spin dynamics we consider $k_{\rm B}T<E_{\sigma}$. In the
stationary limit, one has to solve
$\sum_{\eta'}[P(\eta')\Gamma^{\eta'\to \eta}-
P(\eta)\Gamma^{\eta\to\eta'}]=0$ for the probabilities $P(\eta) \equiv
P(n,s)$. Tunneling is characterized by rates $\Gamma_{\rm
  r,l}^{\eta\to \eta'}$, subject to the selection rules $n'=n\pm 1$,
$s'=s\pm 1$. These contain the non-Fermi liquid correlations and can
be calculated microscopically for finite temperature \cite{bsk01}.
For simplicity, we quote only the expression for $T=0$
\begin{equation}
  \label{eq:6}
 \Gamma_{\rm r,l}^{i\to j}(E_{i\to j})=\gamma_{\rm
  r,l}\sum_{\rho,\sigma\geq 0}
%\left[
a_{\rho}a_{\sigma}\left(\frac{X^{i,j}_{\rho,\sigma}}
{\hbar \omega_{\rm c}}\right)^{\alpha}
\Theta(X^{i,j}_{\rho,\sigma})
%\right]
\,,
\end{equation}
with $X^{i,j}_{\rho,\sigma}=E_{i\to j}-\rho\varepsilon_{\rho}-\sigma
\varepsilon_{\sigma}$, $\alpha=(1/g_{0}-1)/2$,
$a_{\nu}=\Gamma(1/2g_{\nu}+\nu)/\Gamma(1/2g_{\nu})\nu!$  and  
the intrinsic rates
\begin{equation}
   \label{eq:7}
\gamma_{\rm r,l} =
\left(\frac{\varepsilon_{\rho}}{\hbar\omega_{\rm c}}\right)^{1/2g_{\rho}}
\left(\frac{\varepsilon_{\sigma}}{\hbar\omega_{\rm c}}\right)^{1/2g_{\sigma}} 
\frac{\hbar\omega_{\rm c}G_{\rm r,l}}{e^2\Gamma(1+\alpha)}\,,
 \end{equation} 
 with $\omega_{\rm c}$ the cutoff frequency \cite{voit}, $G_{\rm r,l}=(\pi e
 \Delta_{\rm r,l}/\omega_{\rm c})^{2}/h$ the tunneling conductances, and
 $E_{\rm i\to j}$ obtained from the charge and spin addition energies
 corresponding to the states $i$ and $j$. The neutral spin and charge modes
 (energy scales $\varepsilon_{\rho}$ and $\varepsilon_{\sigma}$) are fully
 taken into account. For the spin relaxation, obeying the selection rules
 $s'=s\pm2$ and $n'=n$, we use the detailed-balance result
\begin{equation}
\label{eq:3}
\Gamma_{\rm rel}^{s\to s'}=\left\{
\begin{array}{ll}
w\, & |s'|<|s|\\
w \exp\left[-{1\over 2}\beta E_\sigma (s'^2 - s^2)\right]\, & |s'| > |s|
\end{array}
\right.
\end{equation}
containing a phenomenological rate $w$, and $1/\beta\equiv k_{\rm B}T$.

\begin{figure}[t]%[htbp]
 \setlength{\unitlength}{1cm}
  \includegraphics[clip=true,width=8.2cm,keepaspectratio]{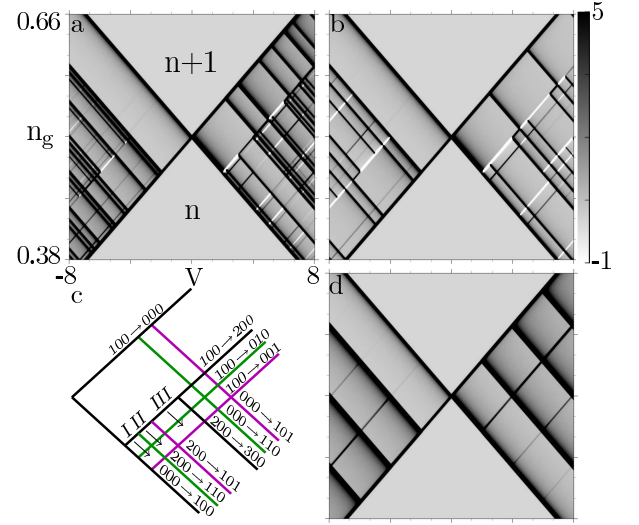}
\vskip-0.1cm 
\caption[]{Differential conductance at $k_{\rm B}T=10^{-2} E_{\sigma}$ 
  in the ($V$,$n_{\rm g})$-plane with asymmetry
  $A=100$, interaction strength in the leads $g_{0}=0.9$, charging
  energy $E_{\rho}=25 E_{\sigma}$, and spin relaxation $w=0$ ($V$ in
  units of $E_{\sigma}/e$); (a) 
  $\varepsilon_{\rho}=3.0 E_{\sigma}$,
  $\varepsilon_{\sigma}=2.2E_{\sigma}$; (b) 
  $\varepsilon_{\rho}=15 E_{\sigma}$,
  $\varepsilon_{\sigma}=2.5E_{\sigma}$; (c) states involved in the
  transitions for $V>0$ near the linear conductance peak, and denoted
  by the triples of integer quantum numbers $(s,\sigma,\rho)$; (d)
  non-interacting case,
  $\varepsilon_{\rho}=\varepsilon_{\sigma}=2E_0$. Right edge: grey
  scale in arbitrary units, labels and scales in a,b,d
  are identical.}
\label{fig:1}
\end{figure}

Numerical results for the differential conductance $G$ in the plane of
bias $V$ and gate voltage $V_{\rm g}\propto n_{\rm g}$ are shown in
Fig.~\ref{fig:1} for $g_{0}=0.9$, $w=0$, $k_{\rm B}T=10^{-2} E_{\sigma}$,
and non-symmetric barriers near a charge transition $n\leftrightarrow
n+1$ ($n$ even). 
The asymmetry $A=G_{\rm l}/G_{\rm r}$ has been
assumed to be large, and the temperature low, in order to emphasize
also weaker features in the conductance. The intersection points at
the centers of the grey-scale panels ($n_{\rm g}=n_{\rm res}$, $V=0$)
denote a linear conductance peak. The grey areas denote the Coulomb blockade
regions. Black and white lines for $V\neq 0$ correspond to positive
and negative differential conductance peaks, respectively. Here,
transitions involving additional spin states as well as neutral spin
or charge {\em modes} in the quantum dot become relevant
(Fig.~\ref{fig:1}c). Remarkably, it is the spin-charge separation in
the dot that induces negative  conductance peaks which
separate certain spin states. Independently of the asymmetry $A$,
these features are {\em not} present when spin-charge separation is
removed, i.e.
$\varepsilon_{\rho}=\varepsilon_{\sigma}=2E_{0}$~(Fig.~\ref{fig:1}d).

In order to understand the physics behind the negative differential
conductances we concentrate in the following on the regions denoted as I, II
and III (Fig.~\ref{fig:1}c). Here, for $k_{\rm B}T<E_{\sigma}$, only the five
states $\eta=(n,s=0), (n+1,s=\pm1),(n,s=\pm 2)$ are necessary to obtain a good
approximation for the conductance. We assume $A>1$ and that the electrons flow
from right to left for $V>0$. Then, the quantum dot with lower particle number
$n$ will have a higher occupation probability. This implies that negative
conductances can only occur parallel to the transition line $(100)\to(000)$, see Fig.~\ref{fig:1}c.

Using this model, the Master equation can be solved analytically using the
rates from Eqs.~(\ref{eq:6})-(\ref{eq:3}) and the energies $E_{0\to 1}=eV/2 +
E_{\rho}(n_{\rm g}-n_{\rm res})$, $E_{2\to 1}=E_{0\to 1}+2E_{\sigma}$ and
$E_{1\to 0}=eV-E_{0\to 1}$, $E_{1\to 2}=eV-E_{2\to 1}$.  The current $I(V)$
can be evaluated, and from that $G=\partial I/\partial V \equiv N/D$. Keeping
only the dominant terms near the transition lines parallel to $(100)\to
(200)$, with $g_0\approx 1$, one obtains $D=[\Gamma_{\rm l}^{1\to
  0}\Gamma_{\rm r}^{2\to 1}+2\Gamma_{\rm r}^{0\to 1}(\Gamma_{\rm r}^{2\to
  1}+\Gamma_{\rm l}^{1\to 2})+w(\Gamma_{\rm l}^{1\to 0}+\Gamma_{\rm l}^{1\to
  2}+2\Gamma_{\rm r}^{0\to 1})]^{2}>0$ and the numerator
\begin{equation}
  \label{eq:8}
  N=e\vartheta\Gamma_{\rm r}^{0\to 1}\sum_{p=\pm 1}
\left[\vartheta K_{1}^{(p)}+\delta K_{2}^{(p)}\right]
\end{equation}
with $\vartheta=2(w+\Gamma_{\rm r}^{2\to 1})$, 
$\delta=\Gamma_{\rm r}^{2\to 1}
-2\Gamma_{\rm r}^{0\to 1}$ and 
\begin{equation}
  \label{eq:9}
K_{1}^{(p)}= \Gamma_{\rm r}^{0\to 1} 
\frac{\partial\Gamma_{\rm l}^{1\to 1+p}}
{\partial V}\,; K_{2}^{(p)}= p\Gamma_{\rm l}^{1\to 1-p}
\frac{\partial\Gamma_{\rm l}^{1\to 1+p}}{\partial V}\,.
\end{equation}
From Eqs. (\ref{eq:8}) and (\ref{eq:9}) one recognizes that for obtaining a
negative differential conductance one needs $N<0$. This implies $\delta
\sum_{p}K_{2}^{(p)}<-\vartheta\sum_{p}K_{1}^{(p)}$ since
$\vartheta\sum_{p}K_{1}^{(p)}$ is always positive.  From Eq.~(\ref{eq:6}) one
finds that the rate changes depend crucially on the presence and strengths of
the correlations via $a_{\nu}$.  Without correlations
($g_0=g_{\rho}=g_{\sigma}=1$), the rates are integer multiples of $\gamma_{\rm
  r,l}$. In particular, one finds $\Gamma_{\rm r}^{2\to 1}=2\Gamma_{\rm
  r}^{0\to 1}$ such that {\em always} $G\ge 0$ (Fig.~\ref{fig:1}d), even in
the most favorable limit $A\to \infty$ and $w=0$. This suggests that it is the
influence of the intra-dot non-Fermi liquid correlations on the rates which
yields the negative differential conductance.

To support this we introduced correlations in the dot ($g_{\sigma}>1$,
$g_{\rho}<1$) while keeping the leads non-interacting ($g_0=1$), for
$w=0$ and $A\to\infty$, with a fixed $G_{\rm r}$. One finds that along
the line $(100)\to(200)$ one has $K_{2}^{(1)}>0$, and
$K_{2}^{(-1)}=0$. This implies {\em always} negative conductance in I
and II since $\delta<0$. The state $(200)$ acts as a bottleneck for
the current, since it cannot relax to the ground state via a single
tunneling event. It accumulates probability at the expense of $(000)$.
In region III, due to the additional activation of the charge wave
state $(101)$ in $\Gamma_{\rm r}^{2\to 1}$, depending on
$a_{\sigma=1}+a_{\rho=1}\lessgtr 1$ one has $\delta\lessgtr 1$ and the
conductance can be both negative and positive. However, for
interactions with $V_{0}>2 V_{\rm x}$ one always gets $G>0$. In
Fig.~\ref{fig:1}b, region III is absent due to $g_{\rho}\ll 1$ (strong
correlation). Therefore, we find only negative conductance as long as
$(300)$ does not contribute.  Similarly, one can analyze the
conductance associated with higher quantum numbers
($s$,$\sigma$,$\rho$). In all of the cases analyzed we have found that
the {\em necessary}
ingredient for negative conductance
is the spin-charge separation.
 \begin{figure}[t]%[htbp]
\setlength{\unitlength}{1cm}\vskip-0.5cm
\includegraphics[clip=true,width=8.2cm]{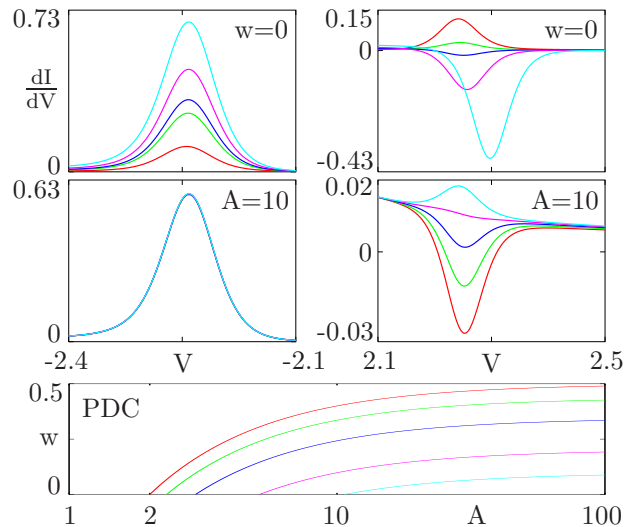}
\vskip-0.1cm
\caption[]{Differential conductance (units $e^2\gamma_{\rm r}/E_{\sigma}$) 
  at positive voltages $V$, in units $E_{\sigma}/e$ (right), near
  $(100)\leftrightarrow (200)$, and at $-V$ (left), for
  $g_{\rho}=0.135$, $g_{\sigma}=1.25$ and $g_{0}=0.9$, at $n_{\rm g}$
  in region I, with $k_{\rm B }T=10^{-2}E_{\sigma}$. Top panels:
  $w=0$, with $A=1$ (red), 2 (green), 2.6 (blue), 5 (magenta) 50
  (cyan).  Middle: $A=10$ with $w/\tilde\gamma=0.3$ (red), 0.35
  (green), 0.4 (blue), 0.45 (magenta), 0.5 (cyan). Bottom: phase
  diagram of the crossover between positive and negative conductance
  at $T=0$ for $g_{0}=1$ (red), 0.9 (green), 0.8 (blue), 0.7
  (magenta), 0.65 (cyan) for $n_{\rm g}$ in region I ($w$ in units $\tilde\gamma =\gamma_{\rm
    l}(E_{\sigma}/\hbar\omega_{\rm c})^{\alpha}$).}
\label{fig:2}
\end{figure}
In order to show that the states with higher total spins are
responsible for the occurrence of negative conductances, we have
studied the influence of spin relaxation with varying strength.
Figure~\ref{fig:2} shows specific numerical results for the
differential conductance at positive and negative biases for $n_{\rm
  g}$ fixed in region I near the $(100)\to (200)$ transition, when the
asymmetry $A$ (top panels) and the relaxation $w$ (middle panels) are
varied.

At $w=0$, for small asymmetries, say $A<2$, all peaks are positive. For larger
asymmetries, $A>2$, the peak corresponding to positive bias can become
negative (top panels). At fixed asymmetry ($A=10$, middle panels), by
increasing the relaxation rate of the state $(200)$ towards the intrinsic rate
$\tilde\gamma =\gamma_{\rm l}(E_{\sigma}/\hbar\omega_{\rm c})^{\alpha}$, the
height of the negative conductance peak is reduced, while the corresponding
positive conductance peak for $V<0$ is unchanged. When $w\ll\tilde\gamma$,
the $(200)$ state is stable and the negative conductance feature is strong.
However, when $w>\tilde\gamma/2$, the $(200)$ state becomes unstable in favor
of $(000)$ and the negative conductance vanishes since the ground state
channels are re-opened.

For finite asymmetry, $A>1$, and with spin relaxation, $w>0$, one
finds a finite critical value, $A_{\rm c}$, for each negative
conductance feature such that one gets $G>0$ for $A<A_{\rm c}$. The
critical trajectories $w_{\rm c}(A)$ for which the negative
conductance peak disappears depends on the state with the highest spin
$s_{\rm max}>1$ involved in the transport, on the interaction strength
in the leads, and on the relaxation. In the above example with $s_{\rm
  max}=2$, $w_{\rm c}(A)$ is obtained by setting $N=0$ in
Eq.~(\ref{eq:8}) at the voltage corresponding to the position of the
negative conductance peak and finite interaction in the leads. This
phase trajectory is shown in Fig.~\ref{fig:2} (bottom) for different
$g_{0}$. The negative conductance depends crucially on the interaction
in the leads $g_{0}$. For decreasing $g_{0}$ tunneling becomes less
efficient, giving spin relaxation increased importance. However, {\em
  with interactions in the leads}, the negative conductance features,
once they are stabilized, can even become enhanced as compared with
their positive counterparts at negative voltages.  Similar results are
obtained for higher $s_{\rm max}$, but spin relaxation is then more
efficient in preventing their formation, $w_{\rm c}(s_{\rm
  max}+1)<w_{\rm c}(s_{\rm max})$.

\begin{figure}[t]%[htbp]
\setlength{\unitlength}{1cm}
\includegraphics[clip=true,width=8.2cm]{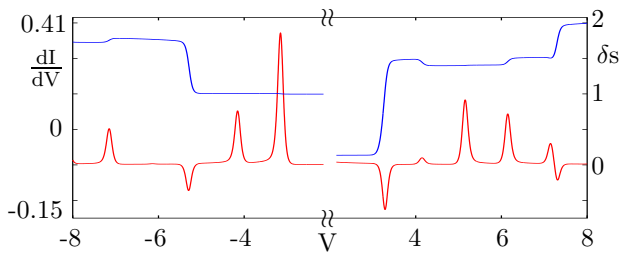}
\vskip-0.1cm 
\caption[]{ Differential conductance 
(red curve, left scale, units $e^2\gamma_{\rm r}/E_{\sigma}$)
  and spin variance (blue curve, right scale) as a function of bias voltage
  $V$ (unit $E_{\sigma}/e$) for $n_{\rm g}$ in region I, $A=100$, $w=0$,
  $k_{\rm B}T=2\cdot 10^{-2}E_{\sigma}$, $E_{\rho}=25 E_{\sigma}$,
  $\varepsilon_{\sigma}=2.5E_{\sigma}$, $\varepsilon_{\rho}=15 E_{\sigma}$,
  $g_{0}=0.9$.}
\label{fig:3}
\end{figure}
Figure~\ref{fig:3} shows the differential conductance and the spin
fluctuations $\delta s\equiv [\sum_{n,s}(s-\langle
s\rangle)^{2}\,P(n,s)]^{1/2}$ as a function of the bias for $g_{0}=0.9$ with
$A=100$. The conductance shows positive and negative peaks for increasing bias
(beyond the first two ground state peaks not shown here). One observes a step
in the spin fluctuation near the negative conductance peaks. This is due to
the participation of the higher-spin states in the transport. For example, the
first negative peak at $V=2.3E_{\sigma}/e$ in Fig.~\ref{fig:3} contains mainly
contributions from $s=2$.  Consequently, the spin variance jumps to a value
dominated by $P(n,s=2)$. This remains dominant even if new conductance
channels enter (subsequent positive peaks). It changes only when the $s=4$
spin channel enters (second negative peak at about $V=7.3E_{\sigma}/e$). Since
we consider the transition $n\leftrightarrow n+1$ with $n$ = even, we conclude
that for positive voltage, {\em even} values of the spin dominate the spin
dynamics.  For negative voltages {\em odd} spins are dominant.  Strong steps
in $\delta s$ are also found at the voltages of the positive conductance peaks
with $n_{\rm g}$ in region III. This indicates that they are an intrinsic
feature of the contribution to transport of the states with higher spins.

In conclusion, we have investigated the non-linear, sequential
transport in a 1D non-Fermi liquid quantum dot embedded in a
correlated electron system including spin. We have found that there
are distinct negative differential conductances induced by spin-charge
separation, if asymmetric tunnel barriers connect the quantum dot to
the leads. We found that one can -- {\em without applying a magnetic
  field} -- stabilize states with higher total spins in certain
regions of the parameter plane spanned by bias and gate voltage if the
asymmetry of the tunnel barriers exceeds a critical value depending on
spin relaxation mechanisms competing with the non-Fermi liquid
correlations. The participation in the transport of the states with
higher total spins is indicated by the sensitivity to spin-flip
relaxation and by strong changes in the spin fluctuations. 
The predicted phenomenon should occur in quasi-1D quantum dots
containing even {\em many} electrons for spin states different from
the ground state but {\em not} necessarily with maximum total spin.
The physical origin of this new correlation-induced trapping
phenomenon are the non-Fermi liquid properties of the system which
lead to spin-charge separation and non-trivially enter the tunneling
rates. The results in Fig.~\ref{fig:2} show that the negative
differential conductances are not restricted to very large asymmetry.
Depending on the interaction parameters, they can occur also for
moderate asymmetries of the order of 2 and even smaller. We expect
that the above non-Fermi liquid phenomenon will be accessible using
quasi-1D electron systems with moderately controllable tunnel contacts
such as carbon nanotubes, and quasi-1D semiconductor quantum wires.
For the latter, asymmetric barriers seem to us to be the genuine case.

Financial support of the EU via RTN FMRX-CT2000-00144 and the Italian MURST,
PRIN02 is acknowledged.

\end{document}